\documentclass[12pt]{iopart}

\usepackage{amscd}
\usepackage{amsfonts}
\usepackage{amssymb}
\usepackage{amsthm}
\usepackage{mathrsfs}
\usepackage{amsfonts}
\usepackage{verbatim}
\usepackage{color}

\begin{document}

\title[Constructing positive maps]{Constructing positive maps from block matrices}

\author{Yu Guo$^{1,2}$ and Heng Fan$^3$}

\address{$^1$~Department of Mathematics, Shanxi Datong University, Datong 037009, China}
\address{$^2$~Institute of Optoelectronic Engineering, Taiyuan University of Technology, Taiyuan 030024, China}
\address{$^3$~Institute of Physics, Chinese Academy of Sciences, Beijing 100190, China}
\ead{yuguophys@yahoo.com.cn; hfan@iphy.ac.cn}
\begin{abstract}
Positive maps are useful for detecting entanglement in quantum
information theory. Any entangled state can be detected by some
positive map. In this paper, the relation between positive block
matrices and  completely positive trace-preserving maps is
characterized. Consequently, a new method for constructing decomposable
maps from positive block matrices
is derived. In addition, a method for constructing
positive but not completely positive
maps from Hermitian block matrices is also obtained.
\end{abstract}

\maketitle

\section{Introduction}

Positive map plays a crucial role in quantum information theory
\cite{Nielsen,Horodecki1996,Lewenstein2001pra}. In the
context of quantum physics, every quantum operation is characterized
by a trace-preserving completely positive map transforming quantum
states to quantum states, where a quantum state is described by a
density matrix, i.e. a positive matrix with unit trace. Positive
but not completely positive (PNCP) maps are useful tools for
detecting entanglement of quantum states \cite{Horodecki1996,Stormer2008}.

Recall that a bipartite
quantum state $\rho\in\mathcal{M}_m\otimes \mathcal{M}_n$ is called
\emph{separable} if it can be written as a convex combination
\cite{Horodecki2001qic,Werner1989}
\begin{eqnarray}
\rho=\sum_ip_i\rho_i^{(1)}\otimes\rho_i^{(2)},\qquad  \sum_ip_i=1,\
p_i\geq0,
\end{eqnarray}
where $\rho_i^{(1)}$ and $\rho_i^{(2)}$ are quantum states in
$\mathcal{M}_m$ (i.e. the algebra of all $m\times m$ complex matrices)
and $\mathcal{M}_n$, respectively. Otherwise, $\rho$ is called
\emph{entangled}. It is well known that a bipartite state
$\rho\in\mathcal{M}_m\otimes \mathcal{M}_n$ is separable if and only
if \cite{Horodecki1996}
\begin{eqnarray}
({\rm id}_m\otimes\Phi)\rho\geq0 \label{1}
\end{eqnarray}
holds for any PNCP map
$\Phi:\mathcal{M}_n\rightarrow\mathcal{M}_m$.
That is, if $({\rm id}_m\otimes\Phi)\rho$ is not positive,
then $\rho$ is entangled; namely,
$\rho$ is detected by $\Phi$.

A well-known example of PNCP map is the transpose $\tau$, another
one is the so-called reduction map defined by $\Phi(A)={\rm
tr}(A)I-A$ \cite{Horodecki1999pra}. The trace map ${\rm tr}(\cdot)$
is one of the most frequently-used completely positive (CP) maps in
quantum literature. Every CP map
$\Phi:\mathcal{M}_m\rightarrow\mathcal{M}_n$ admits the form of
\cite{Choi75,Kraus}
\begin{eqnarray}
\Phi(A)=\sum_iX_iAX_i^\dag,\qquad A\in\mathcal{M}_m,
\end{eqnarray}
where $X_i$ are $n\times m$ matrices. In particular, $\Phi$ is
trace-preserving (i.e. ${\rm tr}(\Phi(A))={\rm tr}(A)$ for any
$A\in\mathcal{M}_m$) if and only if $\sum_iX^\dag X_i=I_m$, and in
such a case, it is called a \emph{quantum channel} in quantum
physics.

The structure of positive maps has been studied extensively by researchers
on both mathematics and quantum physics
\cite{Choi75,Choi75-2,
Choi72,Choi82,Jamiolkowski,Miszczak,Havel,Chruscinski2008,Chruscinski2009pra,
Chruscinski2013jpa,Chruscinski2011jpa,
Tang1986,Osaka1993,Kye2003,
Ha2003,Ha2012jpa,Terhal2001,Benatti,Hou2010jpa,Hou2011jpa,
Robertson,Breuer,Hall2006,Tanahashi}.
Several methods of constructing PNCP maps are proposed
(see \cite{Chruscinski2013jpa,Ha2012jpa} and Refs. therein for detail).
The transpose map and the reduction map are the most
well known decomposable maps;
the first example of an indecomposable map was given by
Choi \cite{Choi75-2}; some variants of Choi type maps are
proposed in \cite{Chruscinski2013jpa,Ha2012jpa};
in \cite{Chruscinski2009pra,Chruscinski2013jpa},
a new type of indecomposable maps are deduced via the reduction map;
the Robertson map \cite{Robertson} and the Breuer map \cite{Breuer,Hall2006}
are indecomposable maps as well;
some other indecomposable maps are introduced
in \cite{Osaka1993,Terhal2001,Tanahashi};
Ha \cite{Ha2003} extended Choi's another example in \cite{Choi72};
the reduction and Robertson maps was generalized by
Chru\'{s}ci\'{n}ski et al.\cite{Chruscinski2011jpa},
where the generalized reduction maps are decomposable
maps and the generalized Robertson maps are indecomposable ones;
in \cite{Hou2010jpa,Hou2011jpa}, the elementary
operators method was proposed; etc.
The main purpose of this
article is to propose a new method for constructing PNCP maps,
from which we can get some useful tools for detecting entanglement in
quantum information theory since any PNCP map
can lead to an entanglement witness via the
Choi-Jamio{\l}kowski
isomorphism \cite{Choi75,Choi82,Jamiolkowski}
and the entanglement witnesses are
observables that allow us to detect entanglement
experimentally.
We stress here that
the method we proposed is far different from the previous ones since
our method is based on the relation between
the positive block matrix and the trace-preserving
positive map (note that the relation here
is different from the Choi-Jamio{\l}kowski
isomorphism, see Remark in section 2).

The paper is organized as follows: in section 2
we characterize the relation between the positive block
matrices and the trace-preserving PNCP maps (lemma 2.1), and reveal
some relation between bipartite states and quantum channels
(corollary 2.2). Several special bipartite states are considered,
which corresponds to special quantum channels. In addition, we
obtain a sufficient condition for a bipartite state to be separable
(corollary 2.3). In section 3, the correspondence between decomposable
maps and non positive partial transpose (NPPT) block
matrices is derived (theorem 3.1).
It is illustrated with several
examples of new decomposable maps. Section 4 is devoted to a another
way of constructing PNCP maps from Hermitian block matrices (theorem
4.1). Finally, we draw some
conclusions and point out some questions for further investigation in section 5.

For clarity, we list the notations and the terminologies in this note.
$A^t$ denotes
the transpose of $A\in\mathcal{M}_m$, and $A^\dag$ stands for the
transpose of the complex conjugate of $A$, i.e. $A^\dag=\bar{A}^t$,
$A$ is
positive (or positive semi-definite), denoted by $A\geq0$, if
$A^\dag=A$ and its eigenvalues are nonnegative. Let
$\mathcal{M}_m^+$ be the positive part of $\mathcal{M}_m$.
$\mathcal{M}_m\otimes \mathcal{M}_n(=\mathcal{M}_m(\mathcal{M}_n))$
is the algebra of all $m\times m$ block matrices with $n\times n$
complex matrices as entries. A linear map
$\Phi:\mathcal{M}_m\rightarrow\mathcal{M}_n$ is \emph{positive}
(resp. \emph{Hermitian}) if
$\Phi(\mathcal{M}_m^+)\subseteq\mathcal{M}_n^+$ (resp. $\Phi(A)$ is
Hermitian for any Hermitian matrix $A\in\mathcal{M}_m$). Let
$E_{ij}$ be the matrix units of the associated matrix algebra.
$\Phi$ is \emph{completely positive} if $({\rm
id}_k\otimes\Phi)(\sum_{i,j=1}^kE_{ij}\otimes
A_{ij})=\sum_{i,j=1}^kE_{ij}\otimes\Phi(A_{ij})$ is positive for any
positive integer $k$ and every $A=\sum_{i,j=1}^kE_{ij}\otimes
A_{ij}\in(\mathcal{M}_k\otimes\mathcal{M}_m)^+$. A positive map
$\Phi$ is \emph{decomposable} if there exist two CP
maps $\Phi_1$ and $\Phi_2$ such that $\Phi=\Phi_1+\Phi_2\circ \tau$,
where $\tau$ denotes the transpose map (i.e. $\tau(A)=A^t$,
$A\in\mathcal{M}_m$). Otherwise, $\Phi$ is defined to be
\emph{indecomposable}. For $A=\sum_{i,j=1}^mE_{ij}\otimes
A_{ij}\in\mathcal{M}_m\otimes \mathcal{M}_n$,
$A^{t_1}=\sum_{i,j=1}^mE_{ij}^t\otimes A_{ij}$ and
$A^{t_2}=\sum_{i,j=1}^mE_{ij}\otimes A_{ij}^t$ are called the
partial transpose of $A$. It is clear that $(A^{t_1})^t=A^{t_2}$,
$(A^{t_2})^t=A^{t_1}$ and $(A^{t_1})^{t_2}=A^{t}$. We call $A$ a
positive partial transpose (PPT) matrix if $A^{t_{1,2}}\geq0.$
If $A\geq0$ with $A^{t_1}\ngeq0$, we call $A$ a NPPT matrix.

\section{CP maps \& positive block matrices}

For $A=\sum_{i,j=1}^mE_{ij}\otimes A_{ij}\in\mathcal{M}_m\otimes
\mathcal{M}_n$, the \emph{reduced matrix} of $A$, denoted by
$A_{1,2}$, is defined by
\begin{eqnarray*}
&A_{1}={\rm tr}_2(A):=({\rm id}_m\otimes {\rm tr})\sum_{i,j=1}^mE_{ij}\otimes A_{ij}
=\sum_{i,j=1}^m{\rm tr}(A_{ij})E_{ij},&\\
&A_{2}={\rm tr}_1(A):=({\rm tr}\otimes {\rm id}_n)\sum_{i,j=1}^mE_{ij}\otimes A_{ij}
=\sum_{i,j=1}^m{\rm tr}(E_{ij})A_{ij}=\sum_iA_{ii},&
\end{eqnarray*}
where ${\rm tr}(\cdot)$ denotes the trace operation. It is obvious
that ${\rm tr}(A)={\rm tr}(A_1)={\rm tr}(A_2)$ and $A_{1,2}\geq0$
whenever $A\geq0$. We start our discussing with the
famous \emph{Schmidt decomposition}
theorem \cite{Schmidt} which reads as:
Let $H_1$ and $H_2$ be two complex Hilbert spaces with $\dim
H_1=m$ and $\dim H_2=n$, and let $|x\rangle$ be a vector (not
necessarily normalized) in $H_1\otimes H_2$, then there exist
orthonormal sets $\{|e_i\rangle\}$ and $\{|f_i\rangle\}$ of $H_1$
and $H_2$ respectively, and positive numbers $\{\lambda_i\}$, such
that
\begin{eqnarray}
|x\rangle=\sum\limits_i^{r(x)}\lambda_i|e_i\rangle|f_i\rangle,
\end{eqnarray}
where $\sum_i\lambda_i^2=\||x\rangle\|$, $\{\lambda_i\}$ are called
the Schmidt coefficients of $|x\rangle$ and $r(x)\leq\min\{m,n\}$ is
the Schmidt number of $|x\rangle$.

The following lemma is necessary.

\smallskip

\noindent{\bf Lemma 2.1} {\it Let
$A\in(\mathcal{M}_m\otimes\mathcal{M}_n)^+$ and
$|x\rangle\in\mathbb{C}^{m}\otimes\mathbb{C}^{m}$ be a vector (not
necessarily normalized) with ${\rm tr}_2(|x\rangle\langle x|)=A_1$,
then there exists a
trace-preserving CP map
$\Lambda:\mathcal{M}_m\rightarrow\mathcal{M}_n$ such that
\begin{eqnarray}
A=({\rm id}_m\otimes \Lambda)|x\rangle\langle x|. \label{lem2.1}
\end{eqnarray}}

\smallskip

\noindent{\bf Proof} \ Let
$A=\sum_{i,j=1}^mE_{ij}\otimes A_{ij}$,
then
$A_1=\sum_{i,j=1}^m{\rm tr}(A_{ij})E_{ij}$.
Let
\begin{eqnarray*}
A_1=\sum\limits_i\lambda_i^2|\psi_i\rangle\langle\psi_i|
\end{eqnarray*}
be the spectral decomposition of $A_1$.
Suppose
\begin{eqnarray*}
|x\rangle=\sum\limits_{i=1}^{r(x)}\lambda_i|\psi_i\rangle|\phi_i\rangle
\end{eqnarray*}
for some orthonormal set $\{|\phi_i\rangle\}$ in $\mathbb{C}^m$ and
let $\{|\psi_i\rangle\}_{i=1}^m$ be an orthonormal basis of
$\mathbb{C}^m$ induced from the eigenvectors of
$A_1$---$\{|\psi_i\rangle\}_{i=1}^r$
(we assume with no loss of generality that ${\rm
rank}(A_1)=r(x)=r$, $1\leq r< m$; the case of
$r=m$ can be discussed similarly). We write
$E'_{ij}=|\psi_i\rangle\langle\psi_j|$ and
$E_{ij}=\sum_{k,l=1}^m\omega_{kl}^{(ij)}E'_{kl}$ for some complex
numbers $\{\omega_{kl}^{(ij)}\}$, then
\begin{eqnarray*}
A&=&\sum\limits_{i,j=1}^m(\sum\limits_{k,l=1}^m\omega_{kl}^{(ij)}E'_{kl})\otimes A_{ij}\\
&=&\sum\limits_{k,l=1}^mE'_{kl}\otimes(\sum\limits_{i,j=1}^m\omega_{kl}^{(ij)}A_{ij})
=\sum\limits_{i,j=1}^mE'_{ij}\otimes A'_{ij},
\end{eqnarray*}
where
$A'_{ij}=\sum_{k,l=1}^m\omega_{ij}^{(kl)}A_{kl}$.
It follows from
\begin{eqnarray*}
A_1=\sum\limits_{i,j=1}^m{\rm tr}(A_{ij}')E_{ij}'
=\sum\limits_{i=1}^r\lambda_i^2E_{ii}'
\end{eqnarray*}
that
\begin{eqnarray*}
{\rm tr}(A_{ij}')=\left\{\begin{array}{cl}0& {\rm when}\ i\neq j,\\
\lambda_i^2&{\rm when}\ i=j,\end{array}\right.
\end{eqnarray*}
and
\begin{eqnarray*}
{\rm tr}(A_{ij}')=0\quad{\rm when}\ i>r\ {\rm or}\ j>r.
\end{eqnarray*}
Since ${\rm tr}(A_{ii}')=0$ implies $A_{ii}'=0$ when $i>r$,
and thus $A_{ij}'=0$ when $i>r$ or $j>r$.
Consequently,
\begin{eqnarray*}
A=\sum\limits_{i,j=1}^rE_{ij}'\otimes A_{ij}'.
\end{eqnarray*}

Let $\{|\phi_i\rangle\}_{i=1}^n$ be an orthonormal basis of
$\mathbb{C}^n$ extended from $\{|\phi_i\rangle\}_{i=1}^r$. We now
define a linear map $\Lambda:\mathcal{M}_m\rightarrow \mathcal{M}_n$
by
\begin{eqnarray}
\Lambda(|\phi_i\rangle\langle\phi_j|)=A^{\prime\prime}_{ij}
=\left\{\begin{array}{cl}\frac{1}{\lambda_i\lambda_j}A'_{ij}& {\rm when}\ 1\leq i,j\leq r,\\
X_{i}&{\rm when}\ r<i=j\leq n,\\
0&{\rm when}\ r<i,j\leq n,\ i\neq j\end{array}\right. \label{8}
\end{eqnarray}
where $X_i$ are positive matrices in $\mathcal{M}_n$ with ${\rm
tr}(X_i)=1$, $r\leq i\leq n$. then $A=({\rm id}_m\otimes
\Lambda)|x\rangle\langle x|$. It remains to show that $\Lambda$ is a
trace-preserving CP map. By the definition above,
it is easy to see that $\Lambda$ preserves the trace. We next show
that it is completely positive. For clarity, we denote by $A'$ and
$A^{\prime\prime}$ the block matrices $\sum_{i,j=1}^rE_{ij}'\otimes
A_{ij}'+\sum_{i>r}E_{ii}'\otimes X_i$ and
$\sum_{i,j=1}^mE_{ij}'\otimes A_{ij}^{\prime\prime}$ respectively
under the basis $\{E_{ij}'\}$ of $\mathcal{M}_r$. Observe that
\begin{eqnarray*}
A^{\prime\prime}
=\left[\begin{array}{cccccc}\frac{1}{\lambda_1}I&&&&&\\
&\ddots&&&&\\
&&\frac{1}{\lambda_r}I&&&\\
&&&I&&\\
&&&&\ddots&\\
&&&&&I\end{array}\right]A'\left[\begin{array}{cccccc}\frac{1}{\lambda_1}I&&&&&\\
&\ddots&&&&\\
&&\frac{1}{\lambda_r}I&&&\\
&&&I&&\\
&&&&\ddots&\\
&&&&&I\end{array}\right],
\end{eqnarray*}
where $I$ is the $n\times n$ unit matrix. Therefore
$A^{\prime\prime}\geq 0$ iff $A'\geq0$, and in turn, iff $A\geq0$, which
implies that $\Lambda$ is a CP map since the
positive block matrix $A^{\prime\prime}$ is the Choi matrix of
$\Lambda$ \cite{Choi75}.\hfill$\square$

\smallskip

The lemma above implies that any positive block matrix can induce
trace-preserving CP maps.

In general, let $A$ be a Hermitian matrix in
$\mathcal{M}_m\otimes\mathcal{M}_n$ and
$|x\rangle\in\mathbb{C}^{m}\otimes\mathbb{C}^{m}$ be a vector (not
necessarily normalized) with $A_1={\rm tr}_2(|x\rangle\langle x|)$, then there exists a
trace-preserving Hermitian map
$\Lambda:\mathcal{M}_m\rightarrow\mathcal{M}_n$ such that
$A=({\rm id}_m\otimes \Lambda)|x\rangle\langle x|$.
(Notice that $\Lambda$ is a Hermitian map if and
only if the Choi matrix of $\Lambda$ is Hermitian \cite{Choi75}.)

\smallskip

\noindent{\bf Remark} \
i) The trace-preserving CP
map satisfying equation~(\ref{lem2.1}) is not unique. If
$|x\rangle=\sum_{i=1}^{r(x)}\lambda_i|\psi_i\rangle|\psi_i\rangle
\in\mathbb{C}^{m}\otimes\mathbb{C}^{m}$ is
its Schmidt decomposition such that
$A_1={\rm tr}_2(|x\rangle\langle x|)$ and $r(x)=m$,
then the trace-preserving CP map $\Lambda$ is unique.
ii) The
equation~(\ref{lem2.1}) is different from the Choi-Jamio{\l}kowski
isomorphism between block matrices and positive maps. Recall that
the Choi-Jami{\l}kowski isomorphism reads as
\cite{Choi75,Choi82,Jamiolkowski}
\begin{eqnarray}
W=({\rm id}_m\otimes \Phi)(\sum_{i,j=1}^mE_{ij}\otimes E_{ij})
=\sum_{i,j=1}^mE_{ij}\otimes \Phi(E_{ij}), \label{remark}
\end{eqnarray}
where $W\in\mathcal{M}_m\otimes\mathcal{M}_n$ and $\Phi$ is a
positive map from $\mathcal{M}_m$ to $\mathcal{M}_n$.
That is, the Choi-Jamio{\l}kowski
isomorphism is established via a positive map acting on a maximally pure entangled state (i.e. $\frac{1}{m}\sum_{i,j=1}^mE_{ij}\otimes E_{ij}$) while Lemma 2.1 (resp. Corollary 2.2 below) is a relation between
any positive block matrix and the corresponding trace-preserving CP map (i.e. quantum channel) associated with the rank-one block matrix (resp. pure state) which is the purification of the reduced matrix.
In addition, the Choi
matrix $W$ is uniquely determined by $\Phi$ and $\Phi$ is unique
when $W$ is fixed. Also note that the map $\Phi$ in equation~(\ref{remark})
is not necessarily trace-preserving. iii) If ${\rm rank}(A_1)=r<m$,
then $A$ can be viewed as a matrix in
$\mathcal{M}_r\otimes\mathcal{M}_n$.

\smallskip

In quantum literature the term \emph{pure state} is sometimes used
for both rank-one density matrix $|\psi\rangle\langle\psi|$ and the
ket $|\psi\rangle$. A pure state $|x\rangle$ is called a
\emph{purification} of a density matrix $\rho$
if $\rho_1={\rm tr}_2(|x\rangle\langle x|)$.

The following are some consequences of Lemma 2.1.

\smallskip

\noindent{\bf Corollary 2.2} {\it Let
$\rho\in\mathcal{M}_m\otimes\mathcal{M}_n$ be a bipartite density
matrix and $|x\rangle\in\mathbb{C}^m\otimes\mathbb{C}^m$ be a
purification of the reduced density matrix $\rho_1$. Then there
exists a quantum channel $\Lambda$ from $\mathcal{M}_m$ to
$\mathcal{M}_n$ such that
\begin{eqnarray}
\rho=({\rm id}_m\otimes \Lambda)|x\rangle\langle x|. \label{10}
\end{eqnarray}}

\smallskip

That is, any bipartite state can arise from a quantum channel acting on
the purification of the reduced state.

Let
$\rho\in\mathcal{M}_m\otimes\mathcal{M}_n$ be a bipartite density
matrix and $\rho_1$ be a pure state. By Corollary 2.2,
$\rho$ is separable, in
particular, $\rho$ is a product state, i.e.
$\rho=\rho_1\otimes\rho_2$.

\smallskip

\noindent{\bf Corollary 2.3} {\it Let
$\rho\in\mathcal{M}_m\otimes\mathcal{M}_m$ be a bipartite density
matrix. If either $m=2$ and ${\rm rank}(\rho_1)\leq3$ or $m=3$ and
${\rm rank}(\rho_1)\leq2$, then $\rho$ is separable iff it is PPT. }

\smallskip

\noindent{\bf Proof} \ If $m=2$ (or $m=3$) and ${\rm
rank}(\rho_1)=r\leq3$ (or $r\leq 2$), then
$\rho=\sum_{i=1}^rE_{ij}'\otimes A_{ij}'$ is in fact a state in
$r\otimes 2$ (or $r\otimes 3$) bipartite system. The theorem is now
clear from the fact that a state in $m\otimes n$ system with
$mn\leq6$ is separable iff it is PPT \cite{Horodecki1996}.
\hfill$\square$

\smallskip

At the end of this section, we list some special bipartite states
which lead to special quantum channels
(see Table~\ref{tab:1}). A quantum channel
$\Lambda:\mathcal{M}_m\rightarrow\mathcal{M}_n$ is
\emph{entanglement breaking} if $({\rm id}_k\otimes\Lambda)\rho$ is
separable for any state $\rho\in\mathcal{M}_k\otimes\mathcal{M}_m$
\cite{Horodecki2003rmp}. It is showed in \cite{Horodecki2003rmp}
that $\Lambda$ is entanglement breaking iff $\Lambda(A)=\sum_k{\rm
tr}(W_kA)\varrho_k$ for any $A\in\mathcal{M}_m$, where each
$\varrho_k$ is a density matrix in $\mathcal{M}_n$, $W_k\geq0$ and
$\sum_kW_k=I_m$. $\Lambda$ is called a \emph{completely contractive}
channel if $\Lambda(A)={\rm tr}(A)\sigma$ holds for any
$A\in\mathcal{M}_m$ for some fixed state $\sigma\in\mathcal{M}_n$
\cite{Karol}. Let $\rho$ be a bipartite density matrix in
$\mathcal{M}_m\otimes\mathcal{M}_n$. If $\rho$ is a
classical-quantum state (i.e. $\rho=\sum_ip_i|i\rangle\langle
i|\otimes \sigma_i$ with $\{|i\rangle\}$ a orthonormal set of
$\mathbb{C}^m$ and $\sigma_i$ are density matrices in
$\mathcal{M}_n$) and ${\rm rank}(\rho_1)=m$, then one can check that
the channel $\Lambda$ in equation~(\ref{10}) is entanglement breaking. In
particular, if $\rho$ is a product state with ${\rm
rank}(\rho_1)=m$, then the channel $\Lambda$ in equation~(\ref{10}) is
completely contractive. If $\rho=|y\rangle\langle y|$ is a pure
state in $\mathcal{M}_m\otimes\mathcal{M}_m$ with $r(y)=m$, then
$\Lambda$ in equation~(\ref{10}) is a unitary operation (channel), i.e.
$\Lambda(A)=UAU^\dag$ for some unitary matrix in $\mathcal{M}_m$.
\begin{table*}
\caption{\label{tab:1}The correspondence
between the state $\rho$ and the channel $\Lambda$.}
\begin{center}
\begin{tabular}{lll}\hline\hline
$\rho$                                                & System                                & $\Lambda$\\ \hline
$\rho=\sum_ip_i|i\rangle\langle
i|\otimes \sigma_i$ with ${\rm rank}(\rho_1)=m$       & $\mathcal{M}_m\otimes\mathcal{M}_n$ & $\Lambda$ is entanglement breaking \\
$\rho=\rho_1\otimes\rho_2$ with ${\rm rank}(\rho_1)=m$& $\mathcal{M}_m\otimes\mathcal{M}_n$ &$\Lambda$ is completely contractive\\
$\rho=|y\rangle\langle y|$ with $r(y)=m$              & $\mathcal{M}_m\otimes\mathcal{M}_m$ & $\Lambda$ is a unitary channel\\
\hline \hline
\end{tabular}
\end{center}
\end{table*}

\section{Decomposable maps derived from NPPT positive block matrices}

For simplicity, we fix some notations. Let $A$ be a positive matrix
in $\mathcal{M}_m\otimes\mathcal{M}_n$. Write
\begin{eqnarray}
A=\sum_{i,j=1}^mE_{ij}\otimes A_{ij}
=\sum_{k,l=1}^n\tilde{A}_{kl}\otimes E_{kl}, \label{3.1}
\end{eqnarray}
where $E_{ij}$ are matrix units in $\mathcal{M}_m$ and $E_{kl}$ are
matrix units in $\mathcal{M}_n$. That is, $A$ can be denoted by
$[A_{ij}]$ or $[\tilde{A}_{kl}]$. It is straightforward that
\begin{eqnarray*}
\tilde{A}_{kl}=[a_{ij}^{(kl)}]\ {\rm iff}\ A_{ij}=[a_{kl}^{(ij)}].
\end{eqnarray*}
Let $A_1={\rm
tr}_2(A)=\sum_{i=1}^r\lambda_i|\psi_i\rangle\langle\psi_i|$ be its
spectral decomposition and $E_{ij}'=|\psi_i\rangle\langle\psi_j|$.
Then $A$ can be represented as $A=\sum_{i,j=1}^rE_{ij}'\otimes
A_{ij}'$, where $A_{ij}'$ are $n\times n$ matrices. Let
$A_{ij}^{\prime\prime}$ defined as in equation~(\ref{8}), it is clear
that $A^{t_2}\geq0$ iff $(A^{\prime\prime})^{t_2}\geq0$,
$A^{\prime\prime}=[A_{ij}^{\prime\prime}]=\sum_{i,j=1}^mE_{ij}'\otimes
A_{ij}^{\prime\prime}$.

The following is the main result of this paper.

\smallskip

\noindent{\bf Theorem 3.1} {\it Let $A$ be a NPPT matrix in
$\mathcal{M}_m\otimes\mathcal{M}_n$. If
$\Phi:\mathcal{M}_m\rightarrow\mathcal{M}_n$ is a linear map
satisfying $\Phi(E_{ij}')=(A_{ij}^{\prime\prime})^t$, where
$E_{ij}'$, $A_{ij}^{\prime\prime}$ are defined as above, then $\Phi$
is a PNCP map, moreover, it is
trace-preserving and decomposable.}

\smallskip

\noindent{\bf Proof} \ Since the Choi matrix of $\Phi$, i.e. $({\rm
id}_m\otimes \Phi)(\sum_{i,j=1}^mE_{ij}\otimes
E_{ij}')=(A^{\prime\prime})^{t_2}$, is not positive, by Theorem 2 in
\cite{Choi75}, $\Phi$ is not completely positive. In order to show
the positivity of $\Phi$, it suffices to prove
\begin{eqnarray*}
\Phi(|w\rangle\langle w|)\geq0
\end{eqnarray*}
holds for any rank-one projection $|w\rangle\langle
w|\in\mathcal{M}_m$. Let $|w\rangle\langle
w|=\sum_{i,j=1}^mt_{ij}E_{ij}'$. We claim that $T=[t_{ij}]$ is a
rank-one projection. In fact, there exists a unitary matrix $U$ such
that $U|\psi_i\rangle=|e_i\rangle$, $i=1$, 2, $\dots$, $m$, where
$\{|\psi_i\rangle\}$ is the orthonormal basis of $\mathbb{C}^m$
derived from the eigenvectors of $A_1$, and $\{|e_i\rangle\}$ is the
standard orthonormal basis of $\mathbb{C}^m$. It turns out that
\begin{eqnarray*}
U|w\rangle\langle w|U^\dag=\sum\limits_{i,j=1}^mt_{ij}UE_{ij}'U^\dag
=\sum\limits_{i,j=1}^mt_{ij}E_{ij}=[t_{ij}],
\end{eqnarray*}
which implies that $T$ is a rank-one projection. Define $\Lambda:
\mathcal{M}_m\rightarrow \mathcal{M}_n$ by
$\Lambda(E_{ij}')=A_{ij}^{\prime\prime}$. By Lemma 2.1, it is
completely positive, and thus positive. Hence
\begin{eqnarray}
\Lambda(|w\rangle\langle w|)=\sum\limits_{i,j=1}^mt_{ij}\Lambda(E_{ij}')
=\sum\limits_{i,j=1}^mt_{ij}A^{\prime\prime}_{ij}
=[{\rm tr}(T^t\tilde{A}^{\prime\prime}_{ij})]
\geq0.
\end{eqnarray}
It follows that
\begin{eqnarray*}
\Phi(|w\rangle\langle w|)&=&\sum\limits_{i,j=1}^mt_{ij}\Phi(E_{ij}')
=\sum\limits_{i,j=1}^mt_{ij}(A^{\prime\prime}_{ij})^t\\
&=&[{\rm tr}(T^t(\tilde{A}^{\prime\prime}_{ij})^t)]
=[{\rm tr}(T\tilde{A}^{\prime\prime}_{ij})]
=\Lambda(|\bar{w}\rangle\langle \bar{w}|)\geq0,
\end{eqnarray*}
that is $\Phi$ is positive. Define
$\Lambda':\mathcal{M}_m\rightarrow\mathcal{M}_n$ by
$\Lambda'(X)=\overline{\Lambda(X)}$, where $\overline{\Lambda(X)}$
denotes the complex conjugate of $\Lambda(X)$. Then $\Lambda'$ is
completely positive. Since $\Phi=\Lambda'\circ \tau$, $\Phi$ is
decomposable. It is easy to see that $\Phi$ is trace-preserving from
the definition. \hfill$\square$

\smallskip

That is, any NPPT block matrix can deduce some
decomposable map(s). We thus
derive a broad class of decomposable maps.
This result provides a new method of constructing
decomposable positive maps
and thus it can
provide new tools for detecting entanglement since
decomposable maps can detect NPPT
entangled states \cite{Lewenstein2001pra}.

We illustrate our results with some well-known NPPT bipartite density
matrices.

\smallskip

\noindent{\bf Example 3.2}
We consider a $3\otimes 3$ density
matrix,
\begin{eqnarray*}
\rho(a)=\frac{1}{21}\left[\begin{array}{ccc|ccc|ccc}
2&0&0&0&2&0&0&0&2\\
0&a&0&0&0&0&0&0&0\\
0&0&5-a&0&0&0&0&0&0\\ \hline
0&0&0&5-a&0&0&0&0&0\\
2&0&0&0&2&0&0&0&2\\
0&0&0&0&0&a&0&0&0\\ \hline
0&0&0&0&0&0&a&0&0\\
0&0&0&0&0&0&0&5-a&0\\
2&0&0&0&2&0&0&0&2
\end{array}\right],
\end{eqnarray*}
where $2\leq a\leq 5$. It is proved in \cite{Horodecki1999} that
$\rho(a)$ is separable iff $2\leq a\leq 3$; is PPT entangled iff
$3<a\leq 4$; is NPPT entangled iff $4<a\leq 5$. One can check
that $\rho(a)$ is NPPT when $0\leq a<1$. By Theorem 3.1,
\begin{eqnarray}
\Phi_1^{a}([a_{ij}])
=2\left[\begin{array}{ccc}
a_{11}&a_{21}&a_{31}\\
a_{12}&a_{22}&a_{32}\\
a_{13}&a_{23}&a_{33}
\end{array}\right]\nonumber\\
~~~~~~+\left[\begin{array}{ccc}
(5-a)a_{22}+aa_{33}&0&0\\
0&aa_{11}+(5-a)a_{33}&0\\
0&0&(5-a)a_{11}+aa_{22}
\end{array}\right]\label{eg1}
\end{eqnarray}
and
\begin{eqnarray}
{\Phi}_2^a([a_{ij}])
=2\left[\begin{array}{ccc}
a_{11}&a_{21}&a_{31}\\
a_{12}&a_{22}&a_{32}\\
a_{13}&a_{23}&a_{33}
\end{array}\right]\nonumber\\
~~~~~~+\left[\begin{array}{ccc}
aa_{22}+(5-a)a_{33}&0&0\\
0&(5-a)a_{11}+aa_{33}+&0\\
0&0&aa_{11}+(5-a)a_{22}
\end{array}\right]\label{eg2}
\end{eqnarray}
are decomposable when $4<
a\leq 5$ or $0\leq a<1$. In fact, $\Phi_{1,2}^a$ is positive iff
$0\leq a\leq5$ and is completely positive iff $2\leq a\leq4$.

\smallskip

\noindent{\bf Example 3.3} We consider the $m\otimes m$ Werner state
\begin{eqnarray}
\omega=\frac{m-x}{m^3-m}I_A\otimes I_B
+\frac{mx-1}{m^3-m}F,\quad x\in[-1,1], \label{p}
\end{eqnarray}
with $F=\sum_{i,j=1}^mE_{ij}\otimes E_{ji}$ being the flip operator.
It is known that \cite{Horodecki2001qic} $\omega$ is separable iff it
is PPT and in turn, iff $0\leq x\leq1$. Take $m=3$, that is
\begin{eqnarray*}
\omega=\left[\begin{array}{ccc|ccc|ccc}
\frac{1+x}{12}&0&0&0&0&0&0&0&0\\
0&\frac{3-x}{24}&0&\frac{3x-1}{24}&0&0&0&0&0\\
0&0&\frac{3-x}{24}&0&0&0&\frac{3x-1}{24}&0&0\\ \hline
0&\frac{3x-1}{24}&0&\frac{3-x}{24}&0&0&0&0&0\\
0&0&0&0&\frac{1+x}{12}&0&0&0&0\\
0&0&0&0&0&\frac{3-x}{24}&0&\frac{3x-1}{24}&0\\ \hline
0&0&\frac{3x-1}{24}&0&0&0&\frac{3-x}{24}&0&0\\
0&0&0&0&0&\frac{3x-1}{24}&0&\frac{3-x}{24}&0\\
0&0&0&0&0&0&0&0&\frac{1+x}{12}
\end{array}\right].
\end{eqnarray*}
It turns out that
\begin{eqnarray*}
\Phi_3^{3,x}(A)=(3x-1)A+(3-x){\rm tr}(A)I,\quad A\in\mathcal{M}_3,
\end{eqnarray*}
is completely positive when $0\leq x\leq 1$, and is
decomposable when $-1\leq x<0$. In general,
\begin{eqnarray}
\Phi_3^{m,x}(A)=(mx-1)A+(m-x){\rm tr}(A)I,\quad A\in\mathcal{M}_m,\label{eg3}
\end{eqnarray}
is decomposable when
$-1\leq x<0$. Interestingly, for the case of $x=-1$, it is the
well-known reduction map. Furthermore, one can check that
$\Phi_3^{m,x}$ is completely positive iff $0\leq x\leq m$ and if
positive iff $-1\leq x\leq m$.

\smallskip

\noindent{\bf Example 3.4}
For the $m\otimes m$ isotropic state
\begin{eqnarray}
\varsigma=\frac{1-y}{m^2}I_m\otimes I_m+yP^+,
\quad -\frac{1}{m^2-1}\leq y\leq 1,
\end{eqnarray}
with $P^+=\frac{1}{m}\sum_{i,j=1}^mE_{ij}\otimes E_{ij}$ is the
so-called maximally entangled state. It is known that $\varsigma$ is
separable iff $y\leq\frac{1}{m+1}$, and in turn, iff it is PPT
\cite{Horodecki1999pra}. For the case of $m=3$,
\begin{eqnarray*}
\varsigma=\\
\frac{1}{9}\left[\begin{array}{ccc|ccc|ccc}
2y+1&0&0&0&3y&0&0&0&3y\\
0&1-y&0&0&0&0&0&0&0\\
0&0&1-y&0&0&0&0&0&0\\ \hline
0&0&0&1-y&0&0&0&0&0\\
3y&0&0&0&2y+1&0&0&0&3y\\
0&0&0&0&0&1-y&0&0&0\\ \hline
0&0&0&0&0&0&1-y&0&0\\
0&0&0&0&0&0&0&1-y&0\\
3y&0&0&0&3y&0&0&0&2y+1
\end{array}\right].
\end{eqnarray*}
It turns out that
\begin{eqnarray*}
\Phi_4^{3,y}(A)=3yA^t+(1-y){\rm tr}(A)I,\quad A\in\mathcal{M}_3,
\end{eqnarray*}
is decomposable when
$\frac{1}{4}<y\leq1$. In general, we have
\begin{eqnarray}
\Phi_4^{m,y}(A)=myA^t+(1-y){\rm tr}(A)I,
\quad A\in\mathcal{M}_m,\label{eg4}
\end{eqnarray}
is decomposable when
$\frac{1}{m+1}<y\leq1$. Especially, if $y=1$, it reduces to the
transpose map; if $y=0$, it reduces to $\Phi(A)={\rm tr}(A)I$; if
$y=\frac{1}{1-m}$, it reduces to $\Phi(A)={\rm tr}(A)I-A^t$. One can
check that $\Phi_4^{m,y}$ is completely positive iff
$-\frac{1}{m-1}\leq y\leq \frac{1}{m+1}$ and is positive iff
$-\frac{1}{m-1}\leq y\leq 1$.

\smallskip

\noindent{\bf Remark} { The  maps $\Phi_1^a$, ${\Phi}_2^a$,
$\Phi_3^{m,x}$ and $\Phi_4^{m,y}$ range from positive but not
completely positive ones to completely positive ones continuously
when the parameter $a$, $x$ and $y$ vary continuously.
Equations~(\ref{eg1}), (\ref{eg2}), (\ref{eg3}) and (\ref{eg4})
propose new types of decomposable maps.

\section{PNCP maps derived from Hermitian block matrices}

We now consider the relation between the PNCP maps and the
Hermitian block matrices. Let
$A\in\mathcal{M}_m\otimes\mathcal{M}_n$, then $A$ can be denoted by
both $[A_{ij}]$ and $[\tilde{A}_{kl}]$ as in equation~(\ref{3.1}). For
any unitary matrix $U\in\mathcal{M}_m$, let
$\tilde{A}_{kl}^U=U\tilde{A}_{kl}U^\dag$. In the following, we write
\begin{eqnarray}
A^U=\sum\limits_{i,j=1}^mE_{ij}\otimes A_{ij}^U
=\sum\limits_{k,l=1}^n\tilde{A}_{kl}^U\otimes E_{kl},
\end{eqnarray}
that is
$A^U=(U\otimes I_n)A(U^\dag\otimes I_n)
=[\tilde{A}_{kl}^U]=[A_{ij}^U]$.

\smallskip

\noindent{\bf Theorem 4.1.} {\it  Let $A$ be a Hermitian matrix in
$\mathcal{M}_m\otimes\mathcal{M}_n$ and $A\ngeq0$. If $A_{ii}\geq0$
and $A_{11}^{U}\geq0$ for any unitary matrix $U$, then
$\Psi(E_{ij})=A_{ij}$ is a PNCP map.}

\smallskip

\noindent{\bf Proof} \ If $A_{ii}\geq0$ and $A_{11}^{U}\geq0$ for
any unitary matrix $U$, then for any rank-one projection
$|w\rangle\langle w|\in\mathcal{M}_m$, writing
\begin{eqnarray*}
|w\rangle\langle w|=\sum\limits_{i,j=1}^mt_{ij}E_{ij},
\end{eqnarray*}
we have
\begin{eqnarray*}
\Psi(|w\rangle\langle w|)=\sum\limits_{i,j=1}^mt_{ij}\Psi(E_{ij})
=\sum\limits_{i,j=1}^mt_{ij}A_{ij}
=[{\rm tr}(T^t\tilde{A}_{ij})]\geq0.
\end{eqnarray*}
The inequality above holds since
\begin{eqnarray*}
[{\rm tr}(T^t\tilde{A}_{ij})]=A_{11}^U
\end{eqnarray*}
for some unitary matrix $U$. That is, $\Psi$ is positive. On the
other hand, $A=({\rm id}_m\otimes \Psi)(\sum_{i,j=1}^mE_{ij}\otimes
E_{ij})\ngeq0$, thus, it is not completely positive. \hfill$\square$
\smallskip

Theorem 4.1 implies that we can derive PNCP
map(s) from any non positive Hermitian matrix $A$
with $A_{ii}\geq0$
and $A_{11}^{U}\geq0$ for any unitary matrix $U$.
That is, we also can turn the task of constructing
PNCP maps into structuring these special class of block matrices.
This method may lead to indecomposable maps.
Note that indecomposable map can detect the
PPT entangled states \cite{Lewenstein2001pra,Stomer1963}
(PPT entangled states contains weaker entanglement
than that of the NPPT entangled states, so
indecomposable map is more stronger than the decomposable one).
From theorem 4.1 one may obtain
new indecomposable maps
which are useful tools for detecting the PPT entangled states.

\section{Conclusions}

We obtained a correspondence between
the positive block matrix and
the trace-preserving CP map, which is
not the case of Choi-Jamio{\l}kowski
isomorphism.
In particular, we showed that any bipartite
state can be produced from a quantum channel acting on
the purification of the reduced state.
Consequently, we derive new methods of
constructing PNCP maps from the block matrices by exploring
the relation between the structure of the block matrix and
the trace-preserving map.
The correspondence between the block matrix
$A$ and the map $\Lambda$ is characterized (see Table~\ref{tab:2}).
\begin{table*}
\caption{\label{tab:2}
The correspondence between the block matrix $A$ and the map $\Lambda$.}
\begin{center}
\begin{tabular}{lll}\hline\hline
$A\geq0$        & & $\Lambda$ is CP \\
$A\ngeq0$, $A$ is PPT         & &$\Lambda$ is decomposable\\
$A\ngeq0$, $A^\dag=A$, $A_{ii}\geq0$
and $A_{11}^{U}\geq0$, $\forall$ $U$      && $\Lambda$ is decomposable or indecomposable\\
\hline \hline
\end{tabular}
\end{center}
\end{table*}
In particular, examples 3.2--3.3 proposed new concrete types of PNCP maps.
Then, by the Choi-Jamio{\l}kowski
isomorphism, we can obtain entanglement witnesses which can detect entanglement physically.
Therefore our results may shed new light on both the quantum entanglement
theory and the positive map theory in operator theory.

Theorem 4.1 leads to interesting questions for further study:
whether or not there exists a matrix $A$ that
satisfies the condition in theorem 4.1?
It is a difficult issue since the arbitrariness of the unitary
matrix $U$. We conjecture that such
matrix exists.
In addition, is the map can be indecomposable?
If this could be, when is it indecomposable?
We will make a
further research in the future.

\ack{The authors are grateful to the referees for their comments to improve this paper.
YG is supported by the China
Postdoctoral Science Foundation funded project (2012M520603),
the National Natural Science Foundation of China
(11171249, 11101250),
the Natural Science Foundation of Shanxi
(2013021001-1, 2012011001-2),
the Research start-up fund for Doctors of Shanxi Datong University
(2011-B-01) and the Scientific and Technological
Innovation Programs of Higher Education Institutions in Shanxi
(20121015). HF is
supported by the `973' program (2010 CB922904).
The authors thank Shunlong Luo for valuable discussions.}

\end{document}